\documentclass[showpacs,preprintnumbers,amsmath,amssymb,twocolumn]{revtex4}

\usepackage{graphicx}
\usepackage{dcolumn}
\usepackage{bm}
\usepackage{graphicx}
\usepackage{amssymb}
\usepackage{amsmath}
\usepackage[serbian,english]{babel}
\usepackage[utf8x]{inputenc}
\usepackage{amsmath}
\usepackage{graphicx}
\usepackage{color}

\usepackage{dcolumn}

\begin{document}

\title{Topological Dependence of Kepler's Third Law for Collisionless
Periodic Three-Body Orbits with Vanishing Angular Momentum and Equal Masses}

\author{V. Dmitra\v sinovi\' c and Milovan \v Suvakov}
\affiliation{Institute of Physics, Belgrade
University, Pregrevica 118, Zemun, \\
P.O.Box 57, 11080 Beograd, Serbia }

\date{\today}

\begin{abstract}
We present results of numerical calculations showing a three-body orbit's period's $T$ dependence 
on its topology. 
This dependence is a simple linear one, when expressed in terms of appropriate variables, suggesting an exact 
mathematical law. This is 
the first known relation between topological and kinematical properties of three-body systems. 
We have used these results to predict the periods of several sets of as yet undiscovered orbits, 
but the relation also indicates that the number of periodic three-body 
orbits with periods shorter than any finite number is countable.
\end{abstract}


\maketitle

\section{Introduction}

There is, at present, no deeper understanding of periodic three-body orbits in Newtonian gravity, 
than the simple change of scale of spatial and temporal coordinates, see Sect. 10 of Ref. \cite{Landau}, 
that can be compared with Kepler's third law for two-body motion, Ref. \cite{Landau}.
Kepler extracted his laws from the astronomical data concerning two-body periodic orbits
collected by Tycho Brahe and his predecessors. 

Unlike a two-body orbit, a periodic three-body orbit is characterized both by its 
kinematic and geometric properties and by its topology, which can be described 
algebraically by a word, or an element $w({\tt a,b,A,B})$ of free group $F_2$ on two letters ${\tt a, b}$ 
(and their inverses ${\tt A=a}^{-1}, {\tt B=b}^{-1}$), see Refs. \cite{Montgomery1998,Suvakov:2013,Suvakov:2014}.
The algorithm used for ``reading'' of words corresponding to periodic orbits is described in the Appendix 
of Ref. \cite{Suvakov:2014}.

Graphically, this amounts to classifying closed curves according to their ``topologies'' in a plane
with two punctures. 
The closed curves are stereographic projections of periodic orbits from the shape-sphere, 
with three punctures - for a detailed explanation, see Refs. \cite{Montgomery1998,Suvakov:2013,Suvakov:2014},
and for graphic illustrations, see the web-site \cite{gallery} - onto a plane with two punctures, the 
puncture at the ``north pole'' having been projected to infinity.

That procedure leads to the aforementioned free group $F_2$ on two letters ${\tt (a,b)}$, where (for 
definiteness) {\tt a} denotes a clockwise ``full turn'' around the right-hand-side puncture, and {\tt b} 
denotes the counter-clockwise full turn around the other puncture in the plane/sphere.
In this way the topology of an orbit can be transformed into an algebraic object that can be 
further manipulated. 

But, even within this particular method of assigning a sequence of symbols to a topology 
there remains an ambiguity, regarding the question which puncture should be taken 
as the ``north pole'' of the stereographic projection, see Appendix \ref{ss:ambiguity}.
The length of the word generally depends on this choice, see Appendix \ref{s:alternative_word}.
We resolve this ambiguity by using the (common) symmetry axis of all presently known collisionless 
zero-angular-momentum periodic orbits to define the ``north pole'', which we call the ``natural'', 
or ``symmetric'' choice, because it leads to equal numbers $n_{\tt a} = n_{\tt b} = \frac12 n_w$ 
of small letters ${\tt a}$ and ${\tt b}$, as well as equal numbers 
$n_{\tt A} = n_{\tt B} = \frac12 {\bar n}_w$ of capital letters ${\tt A}$, or ${\tt B}$.
These relations need not hold with a different choice of ``north pole'', however, e.g. with
cyclically permuted punctures, generally
$n_{1 \tt a} \neq n_{1 \tt b}$ and $n_{1 \tt A} \neq n_{1 \tt B}$,
see Appendix \ref{s:alternative_word}. 
Moreover, the above-described procedure, is not the only way of assigning a sequence of symbols 
to a topology, for an alternative, together with our results expressed in these alternative terms, 
see Appendix \ref{s:alternative}. 

\begin{center}
\begin{figure}[h!]
\includegraphics[width=1.0\columnwidth]{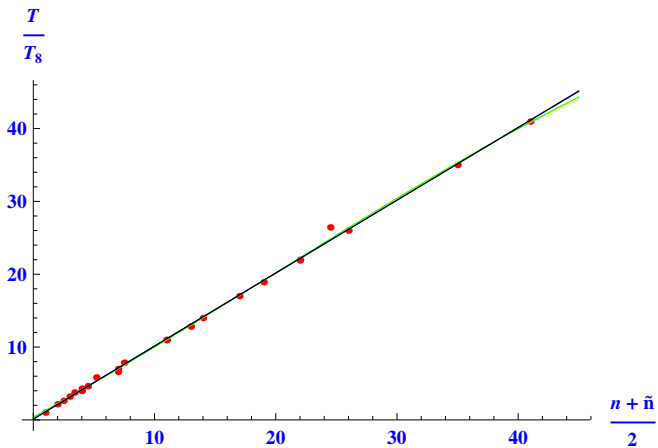}
\caption{(color on line) The rescaled periods $T_r(w)$ of presently known (zero-angular-momentum) 
three-body orbits divided by the period of the figure-8 orbit $T_r(w_8)$, 
versus one half of the length of word $N_w$, i.e., one half of the number of all letters in the free-group 
word $w$ describing the orbit, $N_w/2 = (n_w + {\bar n}_w)/2$, where $n_w$ is the number of small letters 
${\tt a}$, or ${\tt b}$, and ${\bar n}_w$ is the number of capital letters ${\tt A}$, or ${\tt B}$ in the 
letter $w$. Four (linear, quadratic, cubic and quartic) fits are shown as solid lines of different colors, 
yet they overlap so much that the difference can be seen 
only at $N_w > 80$.}
\label{fig:period}
\end{figure}
\end{center}

\begin{center}
\begin{figure}[h!]
\includegraphics[width=1.0\columnwidth]{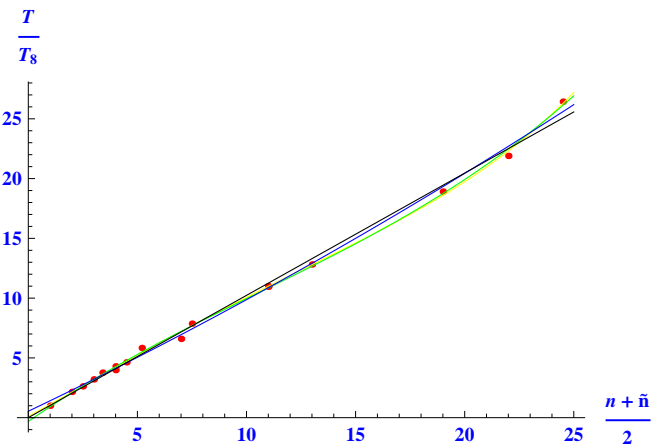}
\caption{(color on line) The rescaled periods $T_r(w)$ of 16 presently known non-satellite
zero-angular-momentum three-body orbits divided by the period of the figure-8 orbit $T_r(w_8)$ 
versus one half of the length of word $N_w$, i.e., 
one half of the number of all letters in the free-group word $w$ describing the orbit, 
$N_w/2 = (n_w + {\bar n}_w)/2$, where $n_w$ is the number of small letters ${\tt a}$, or ${\tt b}$, 
and ${\bar n}_w$ is the number of capital letters ${\tt A}$, or ${\tt B}$ in the letter $w$. 
Four (linear, quadratic, cubic and quartic) fits are shown as solid lines of different colors.} 
\label{fig:period_nonsatellites}
\end{figure}
\end{center}

To date there is no collection of astronomical data regarding periodic orbits of three 
bodies comparable to Brahe's collection of two-body orbits.
Therefore, if one wishes to study general properties of the three-body system one must 
resort to numerical studies. To this end, in this Letter we use the world's total (published) 
data set containing 46 
distinct collisionless periodic orbits,
Refs. \cite{Moore1993,Chenciner2000,Suvakov:2013,Suvakov:2013b,Shibayama,gallery},
to extract the following 
(wholly unexpected)  
linear dependence of the (generalized) Kepler's third law for the ratio $T_r(w)/T_r(w_8)$ of 
``rescaled'' periods (i.e. periods evaluated at one common energy $E$) of three-body orbits,
\begin{equation}
\frac{T_r(w)}{T_r(w_8)} = \frac{T_E(w)|E(w)|^{3/2}}{T_E(w_8)|E(w_8)|^{3/2}} \simeq \frac{N_w}{2} 
= \frac{(n_w + {\bar n}_w)}{2}
\end{equation}
on their topologies $w$, specifically on (one half of) the number of all letters 
$N_w = (n_w + {\bar n}_w)$, see Fig. \ref{fig:period}.
Here $n_w$ is the number of small letters ${\tt a}$, or ${\tt b}$, and ${\bar n}_w$ is the number of 
capital letters ${\tt A}$, or ${\tt B}$ contained in the latter $w$, and $w_8 = {\tt abAB}$ 
if the free-group word describing the figure-8 orbit, Ref. \cite{Montgomery1998,Suvakov:2013,Suvakov:2013b}.
We have divided the total number of letters $N_w$ into two parts
because orbits fall into different classes with distinct values of $n_w$ and ${\bar n}_w$, see 
Table \ref{tab:SD13}.
The worst-case disagreement of this linear dependence with the data is about 10\%,  
though generally it is (much) better, occasionally reaching the limits (six significant 
decimal places) of our numerical precision. In this Letter we point out four more accurate 
(than 10 \%), yet still approximate sub-sequences of orbits, and one possibly exact regularity. 
For clarity's sake, we show in Fig. \ref{fig:period_nonsatellites} the corresponding
graph for non-satellite orbits, only. Note that the range of the abscissa in 
Fig. \ref{fig:period_nonsatellites} only reaches 
the value $N_w = 49$. 

At this point a few words must be said about the statistical significance of results presented in
Fig. \ref{fig:period}, or equivalently about $\#(N_w)$, the number of distinct periodic orbits of 
``length'' $N_w$: 1) At small length values 
one can explicitly count the mathematically allowed orbits and show that many have already been found, 
see Appendix \ref{ss:conjugacy}. 
2) As one increases the number of letters $N_w$, the number of topologically distinct orbits 
$\#(N_w)$ grows rapidly, see Appendix \ref{ss:conjugacy}, 
and the number of presently discovered (and displayed here) orbits pales in comparison with that number.
The number $\#(N_w)$ is not necessarily the same one as the number of physically possible orbits - 
Moore has shown by explicit examples how mathematically allowed orbits disappear as the exponent in 
the potential is reduced from $\rm a = 2$ to $\rm a = 1$ in Newtonian gravity, Ref. \cite{Moore1993}

The large number of still possibly undetected orbits makes the observed linearity of the graph, 
Fig. \ref{fig:period}, at higher values of $N_w$ all the more impressive:
Note that 24, out of grand total of 46 orbits taken from 
Refs. \cite{Moore1993,Chenciner2000,Suvakov:2013,Suvakov:2013b}, extend up to $N_w = 49$. 
These 24 orbits include 10 (non-choreographic) figure-eight satellites from Ref. \cite{Suvakov:2013b}.
Among these 24 there are 16 non-satellite orbits that are shown separately in Fig. \ref{fig:period_nonsatellites}.
The remaining 22 (of 46) orbits are the new ($k$=5, 7, 14, 17, 22, 26, 35, 41 figure-eight satellite) 
choreographies, Ref. \cite{Shibayama}, that extend up to $N_w = 82$ and thus test the proposed linear 
dependence(s) farther into the previously unexplored region. We emphasize that three of these new 
choreographic orbits are not satellites of the figure-eight.

The most precise regularity explains two previously noticed, but unexplained identities:
a) the identity of periods (to 16 decimal places, in Ref. \cite{Simo2002}); and  
b) the identity of actions (to seven significant digits, in Ref. \cite{Galan2002}), 
of two distinct orbits with the same topology, {\it viz.} of Moore's and Simo's figure-8 
solutions, evaluated at equal energies. The same phenomenon was observed among seven independent  
$k=17$ figure-eight satellites reported in 
Ref. \cite{Suvakov:2013b}, and the two butterfly orbits in Ref. \cite{Dmitrasinovic:2014lha}. 
Moreover, the ratio of periods of two different-$k$ satellites equals
the ratio of the corresponding $k$'s, to high precision. 

An extension of this rule to arbitrary $k$ predicts the periods, and actions of new, as yet 
undiscovered three-body orbits and helps one to identify and classify newly discovered orbits, 
as in Ref. \cite{Shibayama}. 
The four less precise linear regularities also predict new families of orbits, with somewhat
lower precision.
Before we proceed to present the evidence and state the regularities in their algebraic forms, 
we must first define what we mean by Kepler's third law in a three-body system?

\subsection*{Kepler's laws for three bodies?}
\label{ss:Kepler}

Kepler's first law, which states that a periodic two-body trajectory must be an ellipse, manifestly 
holds only in the two-body case; his second law, which asserts conservation of angular momentum, 
holds for any number of bodies, but generally does not have a simple geometrical interpretation 
as in the two-body case. His third law, which relates the square of the orbital period $T$ of a 
planet to the cube of the semi-major axis $a$ of its orbit, i.e., $T^2 \propto a^3$, 
also has a proper three-body generalization, despite its appearance.
As only Lagrange's three-body orbits are elliptical, we must address the question what 
corresponds to the semi-major axis $a$ in Kepler's third law for arbitrary periodic three-body orbits? 

The answer to this question can be found by remembering the virial theorem, Ref. \cite{Landau}, 
for the Newtonian potential and then by using the hyper-spherical three-body variables, 
Ref. \cite{Suvakov:2013}. 
As the ``overall size'' of a three-body system is determined by a time-average 
(to be specified below) over one period of the hyper-radius $R$; and 
the virial theorem relates the time-average of the potential energy $V(R) \simeq 1/R$  
to the (constant) total energy $E$, 
so it is the harmonic mean (over one period) of the hyper-radius ${\bar R}$, defined as 
$\frac{1}{\bar R} = \frac{1}{T} \int_{0}^{T} \frac{ d t}{R(t)}$,
that corresponds to the semi-major axis $a$ in the three-body case, see footnote 
\footnote{This is proportional to the time-average of the potential energy over one period
$1/{\bar R} = - \frac{C}{T} \int_{0}^{T} V(r_{i}(t)) d t = - \frac{C}{2}E$, where $E<0$ is the energy
of the system and $C$ is a constant (equal to the integral of the orbit over the shape sphere) 
that depends on the particular orbit:  
every periodic three-body orbit passes through a different sequence of shapes
(including Lagrange's elliptic ones, in which ${\bar R} \propto a$) during one period, 
so each orbit has a different value of the constant $C$, but the validity of the 
scaling law ${\bar R} \propto |E|^{-1}$ should be immediately clear.
This amounts to the well known fact that Kepler's third law follows from 
the spatio-temporal (mechanical) scaling laws, which, in turn, follow from the homogeneity 
of the Newtonian gravity's (two-body) potential, Ref. \cite{Landau}. The scaling laws 
are ${\bf r} \rightarrow \lambda {\bf r}$, $t \rightarrow \lambda^{3/2} t$, and 
consequently ${\bf v} \rightarrow {\bf v} / \sqrt{\lambda}$. The (total) energy
scales as $E \rightarrow \lambda^{-1}E$, the period $T$ as 
$T \rightarrow \lambda^{3/2} T$ and angular momentum as $L \rightarrow \lambda^{1/2} L$.}.

Thus, we (may) replace the ``mean size'' ${\bar R}$ of the three-body system in Kepler's third law 
$T \propto {\bar R}^{3/2}$ with the inverse absolute value of energy $|E|^{-1}$, i.e., $T \propto |E|^{-3/2}$, or 
equivalently $T|E|^{3/2} = {\rm const.}~$ \footnote{The value of the exponent (3/2) follows from 
the (first negative) power of $r$ in the Newtonian two-body potential \cite{Landau}.}. 
The constant on the right-hand-side of this equation is not ``universal'' in the three-body case, 
as it is in the two-body case: it may depend on both the family of the three-body orbit and its 
angular momentum $L$, see Refs. \cite{Henon1976,Henon1977}. 
The angular momentum $L$ transforms differently under scale transformations than the period $T$,
which is the reason why only the vanishing angular momentum $L=0$ is a ``fixed point'' under 
scaling. For this reason, we confine ourselves here to three-body orbits with zero angular momentum, 
and, for simplicity's sake, to equal masses. 

\section{Observed Kepler-like regularities}
\label{s:empirical}

As stated in the Introduction, families of periodic three-body orbits can be 
characterized by the topology of their trajectories in the real configuration space (``braid group''), 
or on the (so-called) shape sphere (``free group''), as described in 
Refs. \cite{Suvakov:2013,Suvakov:2013b,Suvakov:2014,Montgomery1998}, the latter can be specified 
by the conjugacy classes of elements, or ``words'' $w$, for short, consisting 
of letters ${\tt a,b,A=a}^{-1}, {\tt B=b}^{-1}$, that define the free group on two letters ({\tt a, b}). 

Thus, here we must study the dependence of the constant on the right-hand-side of the scaling law
$T(w)|E(w)|^{3/2} = {\rm const}(w)$ 
on the structure of the word $w({\tt a,b,A,B})$ that characterizes a periodic
three-body orbit with zero-angular-momentum.

\subsection*{Numerical observations}
Note that if we classify orbits using only the numbers, 
$n = n_{\tt a} = n_{\tt b}$ and ${\bar n} = n_{\tt A} = n_{\tt B}$, 
of ${\tt a,b}$ and ${\tt A,B}$ letters
respectively, in the orbits' free group group elements, 
then all periodic orbits 
satisfy roughly the linear rule $T \simeq n+{\bar n}$, c.f. column 
$\frac{{\rm T}}{{\rm T}_{\rm M8}}$ 
and $(n+{\bar n})/2$, in Table \ref{tab:SD13} and Fig. \ref{fig:period}. 
\begin{table}[tbh]
\begin{center} 
\caption{Rescaled periods ${\rm T}$ of three-body orbits, their ratios 
with Moore's figure-8 period ${\rm T}_{\rm M8}$, 
and with period ${\rm T}_{\beta}$ of the first orbit $\beta$ in the given section of the Table, 
as functions of the numbers $n_{\tt a}, n_{\tt b}, n_{\tt A}, n_{\tt B}$,
of {\tt a}'s, {\tt b}'s, ${\tt A=a^{-1}}$'s and ${\tt B=b^{-1}}$'s respectively, in the 
free-group word description of the orbit. Note that ``generic'' relations 
$n = n_{\tt a} = n_{\tt b}$ and ${\bar n} = n_{\tt A} = n_{\tt B}$ hold only in the 
``natural'' or symmetric choice of stereographic projection (see the Introduction and Appendix \ref{ss:ambiguity})
and that, due to the time-reversal symmetry of the solutions, the $n$ and ${\bar n}$ may be 
interchanged.}
\begin{tabular}{l@{\hskip 0.05in}c@{\hskip 0.05in}c@{\hskip 0.05in}ccc} \hline \hline 
\setlength
{\rm Label} 
& $\langle {\rm T} \rangle_2$ & $\frac{\langle {\rm T} \rangle_2}{\langle {\rm T}_{\rm M8}\rangle_2}$  
& $\frac{\langle {\rm T} \rangle_2}{\langle {\rm T}_{\beta}\rangle_2}$ & 
$\frac{n+{\bar n}}{n_{\beta}+{\bar n}_{\beta}}$ & $(n, {\bar n})$\\
\hline
\hline
M8 & 26.1281 & 1 & 1 & 1 & 1,1 \\ 
$S8$ & 26.1268 & 0.999951 & 0.999951 & 1 & 1,1 \\ 
\hline
I.B.1 moth I & 68.4636 & 2.62031 & 1 & 1 & 2,3 \\ 
II.B.1 yarn & 205.469 & 7.86391 & 3.00114 & 3 & 6,9 \\ 
\hline
I.A.1 butterfly I & 56.3776 & 2.15774 & 1 & 1 & 2,2 \\ 
I.A.2 butterfly II & 56.3746  & 2.15762 & 0.999944 & 1 & 2,2 \\ 
I.B.5 goggles &  112.129  & 4.29152 & 1.9889 & 2 & 4,4 \\ 
\hline
I.B.7 dragonfly & 104.005 & 3.98059 & 1 & 1 & 4,4 \\ 
I.A.3 bumblebee & 286.192 & 10.9534 & 2.7517 & 11/4 & 11,11 \\
\hline
II.C.2a yin-yang I & 83.7273 & 3.20449 & 1 & 1 & 3,3 \\ 
II.C.2b yin-yang I & 83.7273 & 3.20449 & 1 & 1 & 3,3\\ 
II.C.3a yin-yang II & 334.876 & 12.8167 & 3.9996 & 4 & 12,12  \\ 
II.C.3b yin-yang II & 334.873 & 12.8166 & 3.9996 & 4 & 12,12 \\ 
\hline
I.B.1 moth I & 68.4636 & 2.62031 & 1 & 1 & 2,3 \\ 
I.B.3 butterfly III & 98.4354 & 3.76742 & 1.43778 & 7/5 & 3,4 \\ 
I.B.2 moth II & 121.006  & 4.63126 & 1.76745 & 9/5 & 4,5  \\ 
I.B.4 moth III & 152.33 & 5.83013 & 2.22498 & 11/5 & 5,6 \\ 
I.B.6 butterfly IV & 690.627 & 26.4324 & 10.0875 & 49/5 & 24,25 \\ 
\hline
\hline
\end{tabular}
\label{tab:SD13}
\end{center}
\end{table}
By perusal of Table \ref{tab:SD13} one finds another five sequences of orbits, separated by 
horizontal lines in Table \ref{tab:SD13}. 
The first one of the five sequences in Table \ref{tab:SD13} 
contains two orbits that form in a progenitor-satellite relationship: 
$w({\rm II.B.1~ yarn})=w({\rm I.B.1~ Moth I})^3$, where
the ``II.B.1 yarn'' solution whose topology is described by $w({\rm II.B.1~ yarn})=
{\tt (ba(BAB)ab(ABA)})^3$, 
i.e., as the third power ($k=3$) of the (time-reversed) ``I.B.1 Moth I'' orbit described by 
$w({\rm I.B.1~ Moth I)}={\tt ba(BAB)ab(ABA)}$.
We can see in Table \ref{tab:SD13} that the ratio of their periods equals three, to one part 
in $3 \times 10^4$. This constitutes the presently available data about satellites of orbits 
other than the figure-eight one.
The remaining four sequences in Table \ref{tab:SD13} define ``empirical integer relations/laws'' 
with linearly related periods - we record them here, both as a challenge to mathematicians, and 
as a guide to future numerical searches, as they predict periods of as yet undiscovered orbits.

\subsection*{
Predictions}  

On the basis of these empirical regularities, we predict:
\begin{enumerate}
 \item 
 new ``yin-yang'' orbits with ratios of periods $T({\rm yy k})/T({\rm yy I}) = k =2,3,5,\ldots$; 
 \item 
 new ``butterfly I - goggles'' orbits with ratios of periods $T/T({\rm I.A.1})$
equal to 3,4,5, $\ldots$ ;  
 \item 
 new ``dragonfly - bumblebee'' orbits with $T/T({\rm I.A.3}) = 5/4, 3/2, 7/4, \ldots$;
 \item 
 new ``moth I,II,II - butterfly III'' orbits with $n=6,7, \ldots$. 
The ``butterfly IV'' orbit deviates the most (8\%) from this sequence, and may well 
define a sub-sequence of its own.
\item 
new satellites of ``moth I'', above and beyond the ``yarn'' orbit. 
\end{enumerate}


\section{Three potentially exact propositions}

The above numerical observations led us to search for and examine other examples of integer 
powers of orbits, such as the satellites of the figure-eight solution, Ref. \cite{Suvakov:2013b,Suvakov:2014}.

\subsection*{Statement}

Thus, we found the following regularities/laws in Table \ref{tab:Satellites_mean_reduced} 
that hold to a higher numerical precision than $T \simeq n+{\bar n}$:
\begin{enumerate}
\item All periodic orbits, described by one topology $w$ and normalized to a 
common energy $E$, have almost identical (i.e. valid to a precision comparable to the numerical
precision of the solutions) values of the period $T_1(w) = T_2(w) = T_3(w) = ...$ 
and of the (minimized) action $S_1(w) = S_2(w) = S_3(w) = ...$;

\item Kepler's third law ``constant'' $T(w^k)|E(w^k)|^{3/2}$ of the ``$k$-th satellite orbit''  
(specified by the free-group element $w^k$ where $k$ is an integer) of the ``progenitor 
orbit'' $w$ equals $k$ times the Kepler's third law constant $T(w)|E(w)|^{3/2}$ of the progenitor 
orbit $w$: $T(w^k)|E(w^k)|^{3/2} = k T(w)|E(w)|^{3/2}$.
More simply, periodic orbits normalized to a common energy $E$ have periods related by 
$T(w^{k_{i}}) = k_i T(w)$ \footnote{The numerical precision of this statement in the observed 
cases is generally not as good as that of proposition 1.}. 

\item all periodic orbits described by the topologies $w^{k_{i}}$, where $k_i =1,2,3...$ 
have (minimized) actions $S(w^{k_{i}})$ that are directly proportional 
to the exponent $k_i$: $S(w^{k_{i}})|E(w^{k_{i}})|^{1/2} = k_{i} S(w)|E(w)|^{1/2}$.
\end{enumerate}

Of course, proposition 1. is a special case ($k_i=1$) of propositions 2. and 3.. The fact that
proposition 2. is identical to proposition 3. for Newtonian gravity follows from the 
identity $S = 3 |E| T$, where $S$ is the (minimized) action of a periodic motion, $E$ is its 
energy and $T$ its period. 
In the following we show that proposition 2. is true - by numerical examples; an outline of an 
argument regarding proposition 3., based on Cauchy's residue theorem and the periodicity of three-body motion 
in the case of Newtonian gravity potential, can be found in Ref. \cite{Kepler_chiba2015}.

\subsection*{Numerical evidence}
\label{s:num_evid}

We  use the numerical data from Refs. 
\cite{Suvakov:2013,Suvakov:2013b}: 
periods of orbits (with unit masses $m_i =1$, $i=1,2,3$ and Newtonian coupling constant $G=1$) 
from Ref. \cite{Suvakov:2013} and rescale  it to the common energy $E=-1/2$, which rescaled
periods are tabulated in Table \ref{tab:SD13}. Moreover, we rescale the periods of ``satellites 
of figure-eight'' orbits from Ref. \cite{Suvakov:2013b}
in the same way, and tabulate them in Table \ref{tab:Satellites_mean_reduced} \footnote{For 
a discussion of different ``measurements'' of the periods and their associated errors, see 
Refs. \cite{Supplement,Kepler_chiba2015}}. 

Topologies of ``satellite of the figure-eight'' orbits' are the $k$-th powers of the figure-eight 
orbit (formally, their homotopy class is $({\tt abAB})^k$). 
The initial conditions for Moore's ``figure-eight'' choreography and Sim\'o's figure-eight 
orbit are labelled by $F8$ and $S8$, respectively, in Table \ref{tab:Satellites_mean_reduced}. 
Next note that the ratio of the period $T$ and the period of Moore's figure-8 orbit $T_8$ 
in Table \ref{tab:Satellites_mean_reduced} 
equals the ``slalom/satellite power'' $k=T/T_8$ to (at least) four significant decimal places.
Moreover, as noted by Gal\' an et al. \cite{Galan2002}, the actions of these two orbits are 
also identical, again, up to numerical precision, in accordance with our proposition 1).
Other examples of identity of periods (and actions) of two, or more distinct orbits
from the same topological sector can be found in Table \ref{tab:Satellites_mean_reduced}, 
(see sectors $k=7,11,17$
and comments in Ref. \cite{Suvakov:2013b} about the identity of various orbits)
and in Table \ref{tab:SD13}, see orbits I.A.1 butterfly I and I.A.2 butterfly II. These 
form the numerical evidence for our proposition 1..

Table \ref{tab:Satellites_mean_reduced} also presents the first numerical evidence for propositions 
2. and 3. ``topological scaling law'', albeit limited to the set of satellites of figure-eight solutions.
Propositions 2. and 3. have been confirmed, after the fact, in a new set of 
figure-eight satellite choreographies with $k \in [5, 41]$, 
as described in Ref. \cite{Shibayama}, where they greatly facilitated 
the detection of three choreographic orbits that are not satellites of the figure-eight.
\begin{table}[tbh]
\begin{center} 
\caption{Periods of three-body orbits rescaled to a common energy $E=-1/2$, 
together with the rescaling factor $\lambda = - 2 E$, where $E$ is the energy 
of the orbit before scaling.
T is the period, $T_8$ is the period of Moore's figure-8 orbit and $k$ is the 
``slalom power'' (i.e. $({\tt abAB})^k$ is homotopy class of the orbit). We also list 
Moore's ($M8$) and Sim\'{o}'s ($S8$) figure-eight orbits, for comparison. 
$\langle {\rm T} \rangle$ represents the average value of three ``measurements'' made with
different versions of Runge-Kutta4 and definitions of the period; $\langle {\rm T} \rangle_2$
is the average of the ``best pair'', i.e., eliminating one measurement that deviates the most
from the arithmetic mean; for more on this, see Ref. \cite{Kepler_chiba2015}.}
\begin{tabular}{lc@{\hskip 0.1in}c@{\hskip 0.1in}c@{\hskip 0.1in}c@{\hskip
0.1in}c@{\hskip0.1in}c@{\hskip0.1in}c}
\hline \hline 
\setlength
 \# & $\lambda$ 
 & $\langle {\rm T} \rangle_2$ 
 & $\langle {\rm T} \rangle$ & $\langle {\rm T} \rangle/\langle {\rm T}_{\rm M8}\rangle$ 
 & $\langle T \rangle_2/\langle T_{\rm M8}\rangle_2$ &  ${k}$ \\
\hline
\hline
$M8$ & 2.57428 
& 26.1281 & 26.1284 & 1 & 1 & 1 \\ 
$S8$ & 2.58387 
& 26.1268 & 26.1294  & 1.00004 & 0.99995 & 1 \\ 
\hline
$NC1$ & 3.0086 
& 182.873 & 182.870 & 6.99890 & 6.99910 & 7 \\
$NC2$ & 2.16153 
& 182.873 & 182.871 & 6.99894 & 6.99910 & 7 \\
$O1$ & 3.14090 
& 182.854 & 182.856 & 6.99843 & 6.99837 & 7 \\
$O2$ & 1.99620 
& 182.854 & 182.855 & 6.99839 & 6.99836 & 7 \\
$O3$ & 3.06222 
& 287.333 & 287.334 & 10.9971 & 10.9971 & 11 \\
$O4$ & 2.97791 
& 287.334 & 287.336 & 10.9972 & 10.9971 & 11 \\
$O5$ & 2.98251 
& 365.760 & 365.761 & 13.9987 & 13.9987 & 14 \\
$O6$ & 3.12702 
& 444.154 & 444.155 & 16.9991 & 16.9991 & 17 \\
$O7$ & 3.12618 
& 444.154 & 444.156 & 16.9992 & 16.9991 & 17 \\
$O8$ & 3.10403 
& 444.155 & 444.155 & 16.9991 & 16.9991 & 17 \\
$O9$ & 3.10267 
& 444.155 & 444.156 & 16.9992 & 16.9991 & 17 \\
$O10$ & 3.06755 
& 444.156 & 444.157 & 16.9992 & 16.9992 & 17 \\
$O11$ & 3.06576 
& 444.156 & 444.157 & 16.9992 & 16.9992 & 17 \\
$O12$ & 3.04027 
& 444.157 & 444.158 & 16.9992 & 16.9992 & 17 \\
$O13$ & 2.92754 
& 444.167 & 444.169 & 16.9996 & 16.9996 & 17 \\
$O14$ & 2.89998 
& 444.169 & 444.170 & 16.9997 & 16.9997 & 17 \\
$O15$ & 2.85692 
& 444.169 & 444.170 & 16.9997 & 16.9997 & 17 \\
\hline
\end{tabular}
\label{tab:Satellites_mean_reduced}
\end{center}
\end{table}

\subsection*{Discussion}
\label{ss:discussion}

1) The ratios of the shortest periods $T(\beta)$ in each of the four sequences of orbits 
in Table \ref{tab:SD13} and the figure-eight orbit $T_{\rm M8}$ are all larger than, or equal 
(``dragonfly'') to unity; thus it appears that the figure-8 orbit 
has the shortest scale-invariant period $T|E|^{3/2}$ of all non-colliding zero-angular 
momentum orbits. 
This means that a search for periodic orbits with periods shorter than some finite prescribed 
value can only lead to a finite number of distinct topologies. If we now assume that each
topology has (only) a finite number of satellites for each value of exponent $k$, then 
we are assured that the number of all such orbits must be finite, and that the number of
periodic orbits with arbitrarily long (including infinite, whatever that may mean) periods 
is infinite, but countable/denumerable.

2) The orbits related by the above four ``approximate'' sequences generally fall into different 
``algebraic'' and/or ``geometric classes'' defined in Refs. \cite{Suvakov:2013,Suvakov:2013b}, 
which shows that the topology of an orbit, and its belonging to one of the four sequences,
are not directly related to the classifications based on either of these two concepts. 
Therefore, we conclude that the newly-predicted orbits may fall outside of any previously 
known algebraic and geometric classes and that their initial conditions may lie outside of the 
currently explored subspace of initial conditions. We also note that all of the above orbits 
share a common ``geometrical symmetry'' on the shape sphere: they are all symmetric under 
reflections with respect to the axis passing through the center of the shape sphere and the 
Euler point defining the initial conditions.

3) As noted in Refs. \cite{Suvakov:2013,Suvakov:2014} Montgomery's \cite{Montgomery1998}
prescription for assigning a free-group element conjugacy class to a topology is not unique,
see Appendix \ref{ss:ambiguity}. Different conventions as to the North Pole of the stereographic
projection yield different free-group words, often of different lengths. Thus, 
the total length of a free-group word describing the topology is not a constant,
but generally depend on the convention taken for the stereographic projection from the 
shape-sphere onto the plane with two punctures. 

Here we have used only the ``natural'', or ``symmetric'' convention/rule, i.e. we consistently 
chose the puncture ``opposite'' 
to our initial conditions' Euler point to be the North Pole of the stereographic projection. 
This is a ``natural'' choice because all known zero-angular momentum equal-mass solutions are 
symmetric under reflections about the axis defined by that Euler point and the center of the 
shape sphere, see graphs in Ref. \cite{gallery}. 

This convention yields a unique free-group word for any given orbit with this symmetry, 
with a unique length. It is not clear, however, that all equal-mass zero-angular-momentum 
collisionless periodic orbits have to share this reflection symmetry.
This convention can be extended to the case with two equal masses, because this
symmetry ``survives'' one distinct mass, but not all three.

4) After this Letter was submitted, 
it was pointed out to us that a recent mathematical preprint Ref. \cite{Moeckel:2014}  
contains results similar to ours, {\it viz.} orbits with periods proportional to the number 
of letters in their words, with the {\it caveat} that they hold only for 
certain newly described orbits with non-zero angular momentum, that pass ``very near triple
collisions''. Such orbits have been obtained only formally, i.e. without a computed example,
because triple collisions cannot be regularized. As these orbits do not satisfy two of our 
conditions -  vanishing angular momentum and collisionless orbits, we disregard them in this
context. 

\section{Summary, conclusions and outlook}

In summary, we have analyzed the topological dependence of the scale-invariant period $T_r = T_E|E|^{3/2}$
of presently known periodic three-body orbits and found remarkable (possibly exact) regularities
that can be derived from the corresponding relation, $S(w^k)|E(w^k)|^{1/2} = k S(w)|E(w)|^{1/2}$, 
between the action $S(w^k)$ and energy $E(w^k)$ of the $k$-th satellite orbit described by $w^k$, 
on one hand, and the action $S(w)$, and energy $E(w)$ of the progenitor orbit $w$, on the other. 

The above proposition is, to our knowledge, the first published Kepler-like law for arbitrary
three-body periodic orbits. Besides this one, possibly exact 
new law, there is empirical evidence for other approximate, but still unexplained regularities.

The results 
presented here 
are striking and unexpected. If their validity persists after further improvements of numerical 
precision, they will be signs of deeper mathematical truths.
But, we must at all times keep in mind the fact that these are the results of numerical experiments, 
which are just what their name implies: experiments. 
So, it is clear that numerical experiments cannot be used to prove theorems; 
that is left to mathematicians.

From the practical point of view, perhaps the most important results are those that point towards 
new, as yet undiscovered periodic solutions. 
Our results do not predict the initial conditions of the new orbits, but only their periods,
at given energy.

From the conceptual point of view, perhaps the most important result of this study is that it 
points towards a denumerable, and perhaps even finite number of periodic orbits with periods 
shorter than some pre-specified finite value.  
If the latter (finite number) proposition is true, that would (greatly) constrain the number of 
possible periodic orbits, 
and thus would imply that one could meaningfully search for and perhaps even find {\it all} 
periodic orbits with periods shorter than some prescribed value. At any rate, our results 
indicate that periodic orbits do not form a dense set in the set of all three-body orbits.

These results also pose several questions in physics and astrophysics: 
1) astrophysics: how can these three-body systems be formed in an astrophysical process?
One mechanism for the formation of the figure-eight orbit was suggested by Heggie, 
Ref. \cite{Heggie:2000ad}, and it can be tried for other orbits.
Before one attempts answering such a question, however, one must establish which of the  
new orbits are stable, as only stable orbits have non-vanishing probabilities of being formed.
2) physics: what happens when one replaces the Newtonian gravity with Coulombian electrostatics? 
In that case one ends up with new free parameters (ratios of electric charges), as these coupling 
constants do not dependent on the masses. Do our regularities survive in the case of electrostatics,
and under which conditions? 

We conclude by noting that our results also present new mathematical questions, such as: 
1) how many degenerate minima of the action (integral) are there in each topological sector?
This particular question falls into the realm of the variational calculus in the large, or 
Morse theory, see \cite{Dubrovin}; 
2) how many algebraically distinct conjugacy classes of $F_2$ correspond to any given topology of a
three-body orbit? This ``automorphism'' question is an open problem in combinatorial group theory, 
see Appendix \ref{ss:ambiguity} and Ref. \cite{Vogtmann2002}.
3) how many topologically distinct orbits/algebraically distinct conjugacy classes are there 
as a function of the word length $N_w$ (see Appendix \ref{ss:conjugacy})? This is a 
combinatorial problem.
These and other questions remain as a challenge to mathematicians.

\noindent{\bf Acknowledgments}
This work was supported by the Serbian Ministry of Education, Science and Technological 
Development under grant numbers OI 171037 and III 41011. 
We thank Dr. Mitsuru Shibayama (Osaka U.) for providing one set of measurements of periods 
and for a thoughtful reading of the manuscript, and 
Dr. Ayumu Sugita (Osaka City U.) for alerting us to the discrepancy in the measured value 
of figure-eight orbit's period and for letting us use his code to measure other periods used 
in this paper. Last, but not least we thank Aleksandar Bojarov for his help with generating
data in Table \ref{tab:SD13syzygy3}.

\appendix

\section{Counting topologically distinct orbits}
\label{s:Counting}

\subsection{Ambiguities in algebraic description of topologies}
\label{ss:ambiguity}

The ambiguity mentioned in the Introduction is best illustrated by a simple example,
already shown in Refs. \cite{Suvakov:2013,Suvakov:2014}: the Broucke-Hadjidemetriou-H\' enon (BHH)
family of orbits. This family is described by a simple loop around any one of three poles/punctures
in the shape-sphere. By applying the stereographic projection of the shape sphere onto a plane with two
punctures, a single loop can be ``translated'' into three different ``words'', as can be seen below.

1) If one chooses as the ``north pole'' of the stereographic projection either one of the two 
punctures that are not encircled by the trajectory, then the orbit is denoted by the free-group letters
({\tt a}, or {\tt B}). 

2) If, on the other hand, one chooses the same puncture that is being encircled as the ``north pole'',
that corresponds to {\tt aB}, i.e., a loop around both punctures in the plane.

But a single loop around any one of the three punctures on the original shape-sphere
must be equivalent to a loop around either of the two remaining punctures; consequently, 
{\tt aB} is equivalent to {\tt a} and/or {\tt B}. 

Of course, the time-reversed solutions are
topologically equivalent to the original orbits, i.e.,  ${\tt a} \equiv {\tt A}$,
${\tt B} \equiv {\tt b}$, ${\tt aB} \equiv {\tt bA}$. 
Thus, we see that (up to six) different conjugacy classes of possibly 
different lengths can be assigned to one and the same topology.

The length of a free-group word is constant under such ``changes of punctures'' 
for orbits that are invariant under cyclic permutations of three punctures on the shape sphere.
So far, there are only four known examples of orbits with a constant free-group word length: 
one of them is the figure-eight orbit (and its satellites); the remaining three are the 
new non-figure-eight choreographies found in Ref. \cite{Shibayama}.
It is an open question if any periodic three-body equal-mass orbits exist 
that have no remnant of the full $S_3$ permutation symmetry group?


The transformations/mappings of words defined by rules such as those shown above
are known in mathematics as automorphisms of a free group - in this regard we can do no 
better than to quote Vogtmann: ``The study of automorphism groups of free groups in itself 
is decidedly not new; these groups are basic objects in the field of combinatorial group theory, 
and have been studied since the very beginnings of the subject. Fundamental contributions
were made by Jakob Nielsen starting in 1915 and by J.H.C. Whitehead in the 1930's to 1950's. 
... these groups have attracted a 
tremendous amount of research work in spite of gaps of decades in the sequences of 
papers that deal with them and, in spite of the work that was done
there was still much more unknown about these groups than known.''
Ref. \cite{Vogtmann2002}.
In summary, the effects of automorphisms on conjugacy classes, and in particular 
on the number of classes equivalent under such automorphisms, are not known in general.
But, computations of conjugacy classes of specific words can be readily done 
using an electronic computer, see Appendix \ref{s:alternative_word}.
One can also count the number of conjugacy classes of $F_2$ explicitly, ``by hand'', 
for small word lengths, and/or try and establish upper and lower bounds on this number, 
as in Appendix \ref{ss:conjugacy}. 


\subsection{Counting conjugacy classes of free group elements}
\label{ss:conjugacy}

We are indebted to referee C for raising this question.

1) The number of distinct $N$-letter words in alphabets with four letters is $4^{N}$;
the number of conjugacy classes equals $4^{N}/N$. 

2) The above estimate of $4^{N}/N$ is a gross overestimate of an upper bound on the number of conjugacy 
classes in the free group $F_2$, because one did not take into account the reduction in length 
of all words containing two neighbouring identical small and large letters, e.g. ({\tt a} and 
{\tt A}), or ({\tt b} and {\tt B}). 
Such words immediately boil down to words with $(N-2)$ letters, etc. This reduces the number 
of conjugacy classes of $F_2$ (far) below $4^{N}/N$!

3) Just how much smaller is the correct number than $4^{N}/N$? We cannot say precisely, but we 
have a lower bound instead: the number of conjugacy classes can not be smaller than 
$(2^{N}/N) \times {\rm const}.$. Taking 
only two small, or two capital letters, or one small and a different capital letter, such as 
({\tt a} and {\tt B})  into consideration, the two can not cancel; of course there is more 
than one such choice, and that multiplicity is covered by the constant ${\rm const}.$)! 
This constant is larger than 2, but less than 8, so it makes a difference only at small $N$. 
At $N \leq 10$, this number is generally around the square root of the upper bound. 

4) Thus we have an upper and a lower bound both of which grow as $e^{{\rm d} \times N}/N$, as $N \to \infty$, 
albeit with different values of constant d: for the upper bound ${\rm d} = \log 4$ and for the lower bound 
${\rm d} = \log 2$. It seems natural to assume that the actual number of conjugacy classes also follows 
the $e^{{\rm d_{ex.}} \times N}/N$ asymptotic law, with a exact value of ${\rm d_{ex.}}$ lying between the 
above two values: $\log 2 < {\rm d_{ex.}} < \log 4$.

5) None of the above bounds and estimates took into account the automorphism ambiguity
described in Appendix \ref{ss:ambiguity} above, which means that the 
number of distinct topologies can only be smaller than these estimates.

6) There are further constraints, stemming from physical symmetries and dynamical considerations, 
which must be imposed on the number of conjugacy classes of $F_2$ that are equivalent under 
automorphisms, in order for that number to be equal to the number of distinct orbits: 

a) Time-reversal invariance demands that an orbit and its 
time-reversed version (``inverse'') be viewed as equivalent. 
Needless to say, conjugacy classes of an (non-trivial) element
and its inverse are algebraically different. Therefore, the above number 
of conjugacy classes must be halved in order to obtain the number of distinct orbits.

b) Not all topologies allow zero angular momentum orbits without collisions.
The simplest example is the BHH family, which at zero angular momentum
leads to the v. Schubart orbit, which is a collinear, therefore colliding,
periodic orbit.

The exact computation of the number of topologically distinct orbits
as a function of word length $N_w$ is a difficult combinatorial problem
that remains as a challenge to mathematicians. 

\section{Alternative symbolic sequences}
\label{a:alternatives}

\subsection{Alternative word readings}
\label{s:alternative_word}

We can readily implement cyclic permutations of three punctures in our free-group
word-reading algorithm, see the Appendix in Ref. \cite{Suvakov:2014}, and this results
in the new word lengths shown in Table \ref{tab:SD13syzygy3}.
\begin{table}[tbh]
\begin{center} 
\caption{Rescaled ratios of periods ${\rm T}$ of three-body orbits and Moore's figure-8 period 
${\rm T}_{\rm M8}$, and ratios with period ${\rm T}_{\beta}$ of the first orbit $\beta$ in the 
given section of the Table, as functions of the numbers $n_{\tt a}, n_{\tt b}, n_{\tt A}, n_{\tt B}$,
of {\tt a}'s, {\tt b}'s, ${\tt A=a^{-1}}$'s and ${\tt B=b^{-1}}$'s respectively, in the 
free-group word description of the orbit for both the ``natural choice'' and cyclic permutations of
punctures on the shape sphere. Note that the ``generic'' relations 
$n = n_{\tt a} = n_{\tt b}$ and ${\bar n} = n_{\tt A} = n_{\tt B}$ which hold with the ``natural''
or ``symmetric'' choice of punctures, do not hold with cyclically permuted punctures, i.e., 
$n_{1 \tt a} \neq n_{1 \tt b}$ and $n_{1 \tt A} \neq n_{1 \tt B}$. 
Here we define $n_1 = \frac12 (n_{1 \tt a} + n_{1 \tt b})$ and 
${\bar n}_1 = \frac12 (n_{1 \tt A} + n_{1 \tt B})$. The $n_1$ and ${\bar n}_1$ may still be interchanged.} 
\begin{tabular}{l@{\hskip 0.05in}c@{\hskip 0.05in}c@{\hskip 0.05in}cccc} \hline \hline 
\setlength
{\rm Label} 
& $\frac{\langle {\rm T} \rangle_2}{\langle {\rm T}_{\rm M8}\rangle_2}$  
& $\frac{\langle {\rm T} \rangle_2}{\langle {\rm T}_{\beta}\rangle_2}$ & 
$\frac{n+{\bar n}}{n_{\beta}+{\bar n}_{\beta}}$ & $(n, {\bar n})$ & 
$\frac{n_1 +{\bar n}_1 }{n_{1 \beta}+{\bar n}_{1 \beta}}$ & $(n_1, {\bar n}_1)$  \\
\hline
\hline
M8 & 1 & 1 & 1 & (1,1) & 1 & (1,1) \\ 
$S8$ & 0.999951 & 0.999951 & 1 & (1,1) & 1 & (1,1) \\ 
\hline
I.B.1 
& 2.62031 & 1 & 1 & (2,3) & 1 & (2.5,3) \\ 
II.B.1 
& 7.86391 & 3.00114 & 3 & (6,9) & 3 & (7.5,9) \\ 
\hline
I.A.1 
& 2.15774 & 1 & 1 & (2,2) & 1 & (2.5,2.5) \\ 
I.A.2 
& 2.15762 & 0.999944 & 1 & (2,2) & 1 & (2.5,2.5) \\ 
I.B.5 
& 4.29152 & 1.9889 & 2 & (4,4) & 2 & (5,5) \\ 
\hline
I.B.7 
& 3.98059 & 1 & 1 & (4,4) & 1 & (4,4) \\ 
I.A.3 
& 10.9534 & 2.7517 & 11/4 & (11,11) & 11/4 & (11,11) \\
\hline
II.C.2 
& 3.20449 & 1 & 1 & (3,3) & 1 & (3.5,3.5) \\ 
II.C.3 
& 12.8167 & 3.9996 & 4 & (12,12) & 4 & (14,14) \\ 
\hline
I.B.1 
& 2.62031 & 1 & 1 & (2,3) & 1 & (2.5,3) \\ 
I.B.3 
& 3.76742 & 1.43778 & 7/5 & (3,4) & 17/11 & (4,4.5) \\ 
I.B.2 
& 4.63126 & 1.76745 & 9/5 & (4,5) & 19/11 & (4.5,5) \\ 
I.B.4 
& 5.83013 & 2.22498 & 13/5 & (6,7) & 25/11 & (6,6.5) \\ 
I.B.6 
& 26.4324 & 10.0875 & 49/5 & (24,25) & 123/11 & (30.5,31) \\ 
\hline
\hline
\end{tabular}
\label{tab:SD13syzygy3}
\end{center}
\end{table}
Note that many of the ratios $\frac{n_1 +{\bar n}_1 }{n_{1 \beta}+{\bar n}_{1 \beta}}$ and 
$\frac{n+{\bar n}}{n_{\beta}+{\bar n}_{\beta}}$ in Table \ref{tab:SD13syzygy3} are identical, 
but just as many are not: in particular 
no ratio in the ``moth I - butterfly IV'' sub-sequence is independent of the choice of puncture. 
Perhaps more importantly, the ratios of 
$\frac{n_1 +{\bar n}_1 }{n_{1 \rm M_8}+{\bar n}_{1 \rm M_8}}$, which determine the ratios of 
periods $\frac{{\rm T}}{{\rm T}_{\rm M8}}$ , i.e., the slopes of linear sub-sequences are often 
not-as-good as with the ``natural choice'' $\frac{n+{\bar n}}{n_{\rm M_8}+{\bar n}_{\rm M_8}}$: 
for example, for ``moth I'' orbit, the measured value $\frac{{\rm T}}{{\rm T}_{\rm M8}}=2.62031$ is 
closer to $\frac{n+{\bar n}}{n_{\rm M_8}+{\bar n}_{\rm M_8}} = 2.5$ than to 
$\frac{n_1 +{\bar n}_1 }{n_{1 \rm M_8}+{\bar n}_{1 \rm M_8}} = 2.75$. Similarly, 
for the I.A.1 ``butterfly I'' orbit, the measured value $\frac{{\rm T}}{{\rm T}_{\rm M8}} = 2.15774$
is closer to $\frac{n+{\bar n}}{n_{\rm M_8}+{\bar n}_{\rm M_8}} = 2$ than to 
$\frac{n_1 +{\bar n}_1 }{n_{1 \rm M_8}+{\bar n}_{1 \rm M_8}} = 2.5$, and 
for the II.C.2 ``yin-yang I'' orbit, the measured value $\frac{{\rm T}}{{\rm T}_{\rm M8}} = 3.20449$
is closer to $\frac{n+{\bar n}}{n_{\rm M_8}+{\bar n}_{\rm M_8}} = 3$ than to 
$\frac{n_1 +{\bar n}_1 }{n_{1 \rm M_8}+{\bar n}_{1 \rm M_8}} = 3.5$. The only $\beta$-orbit
whose measured value $\frac{{\rm T}}{{\rm T}_{\rm M8}} = 3.98059$ is not changed by a change 
of word-reading puncture is the I.B.7 ``dragonfly'', as in both the ``natural'' and ``permuted'' cases 
$\frac{n+{\bar n}}{n_{\rm M_8}+{\bar n}_{\rm M_8}} = \frac{n_1 +{\bar n}_1 }{n_{1 \rm M_8}+{\bar n}_{1 \rm M_8}} 
= 4$ holds.
Note also that for the I.B.6 ``butterfly IV'' orbit $\frac{{\rm T}}{{\rm T}_{\rm M8}} = 26.4324$, which 
set the largest deviation (7.8\%) with $\frac{n+{\bar n}}{n_{\rm M_8}+{\bar n}_{\rm M_8}} = 49/2 = 24.5$, 
this deviation becomes twice as large with  
$\frac{n_1 +{\bar n}_1 }{n_{1 \rm M_8}+{\bar n}_{1 \rm M_8}} = 123/4 = 30.75$.
\begin{center}
\begin{figure}[h!]
\includegraphics[width=1.0\columnwidth]{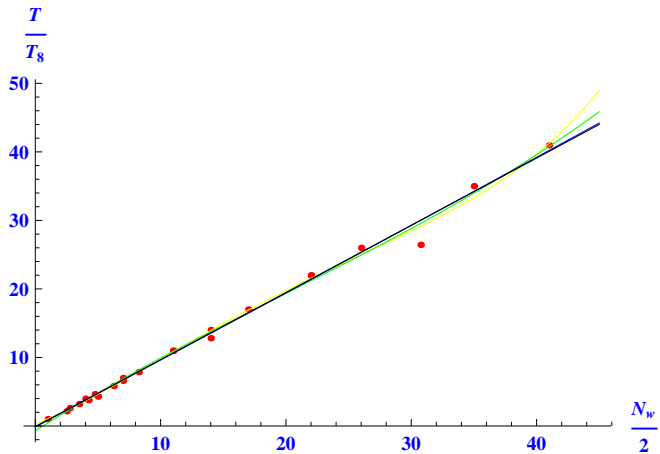}
\caption{(color on line) The rescaled periods $T_r(w)$ of presently known (zero-angular-momentum) 
three-body orbits divided by the period of the figure-8 orbit $T_r(w_8)$ versus one half the word length 
$N_{1w}/2$ in terms of ``permuted puncture word reading''.
Four (linear, quadratic, cubic and quartic) fits are shown as solid lines of different colors.} 
\label{fig:period_syzygy}
\end{figure}
\end{center}

\subsection{Three digit (syzygies) sequences}
\label{s:alternative}

There is an alternative method of assigning a sequence of three symbols 
to any given ``word'' in $F_2$. The rule  for converting ``words'' written in terms of letters 
{\tt a, b, A, B} to ``numbers'' in terms of three digits - (1, 2, 3) - 
is simple: 
Make the substitution {\tt a = 12, A = 21, b = 32, B} = 23 together with 
``cancelling in pairs rule'': 11 = 22 = 33 = (empty sequence). So, for example 
{\tt abAB} = (12)(32)(21)(23) = 12322123 = 123123 for the figure-eight, i.e. a
an increase from $N_w = 4$ to $N_s = 6$ symbols.
The number of symbols $N_s$ changes to those shown in Table \ref{tab:SD13syzygy2}.

\begin{table}[tbh]
\begin{center} 
\caption{Rescaled ratios of periods ${\rm T}$ of three-body orbits and Moore's figure-8 
period ${\rm T}_{\rm M8}$, 
and ratios with period ${\rm T}_{\beta}$ of the first orbit $\beta$ in the given section 
of the Table, as functions of the numbers $n_{\tt a}, n_{\tt b}, n_{\tt A}, n_{\tt B}$,
of {\tt a}'s, {\tt b}'s, ${\tt A=a^{-1}}$'s and ${\tt B=b^{-1}}$'s respectively, in the 
free-group word description of the orbit, as well as of the total length $N_s$ of the 
sequence of integers 1,2,3 used to describe the topology, see text. Subscript $\beta$ denotes the 
first, shortest orbits in a particular sequence of orbits.}
\begin{tabular}{l@{\hskip 0.05in}c@{\hskip 0.05in}c@{\hskip 0.05in}cccc} \hline \hline 
\setlength
{\rm Label} 
& $\frac{\langle {\rm T} \rangle_2}{\langle {\rm T}_{\rm M8}\rangle_2}$  
& $\frac{\langle {\rm T} \rangle_2}{\langle {\rm T}_{\beta}\rangle_2}$ & 
$\frac{n+{\bar n}}{n_{\beta}+{\bar n}_{\beta}}$ & $(n, {\bar n})$ & 
$\frac{N_s}{N_s(\beta)}$ & $N_s$ \\
\hline
\hline
M8 & 1 & 1 & 1 & 1,1 & 1 & 6 \\ 
$S8$ & 0.999951 & 0.999951 & 1 & 1,1 & 1 & 6 \\ 
\hline
I.B.1 moth I & 2.62031 & 1 & 1 & 2,3 & 1 & 18 \\ 
II.B.1 yarn & 7.86391 & 3.00114 & 3 & 6,9 & 3 & 54 \\ 
\hline
I.A.1 butterfly I & 2.15774 & 1 & 1 & 2,2 & 1 & 14 \\ 
I.A.2 butterfly II & 2.15762 & 0.999944 & 1 & 2,2 & 1 & 14 \\ 
I.B.5 goggles & 4.29152 & 1.9889 & 2 & 4,4 & 2 & 28 \\ 
\hline
I.B.7 dragonfly & 3.98059 & 1 & 1 & 4,4 & 1 & 24 \\ 
I.A.3 bumblebee & 10.9534 & 2.7517 & 11/4 & 11,11 & 11/4 & 66 \\
\hline
II.C.2 yin-yang I & 3.20449 & 1 & 1 & 3,3 & 1 & 20 \\ 
II.C.3 yin-yang II & 12.8167 & 3.9996 & 4 & 12,12 & 44/10 & 88 \\ 
\hline
I.B.1 moth I & 2.62031 & 1 & 1 & 2,3 & 1 & 18 \\ 
I.B.3 butterfly III & 3.76742 & 1.43778 & 7/5 & 3,4 & 4/3 & 24 \\ 
I.B.2 moth II & 4.63126 & 1.76745 & 9/5 & 4,5 & 14/9 & 28 \\ 
I.B.4 moth III & 5.83013 & 2.22498 & 11/5 & 5,6 & 22/9 & 44 \\ 
I.B.6 butterfly IV & 26.4324 & 10.0875 & 49/5 & 24,25 & 89/9 & 178 \\ 
\hline
\hline
\end{tabular}
\label{tab:SD13syzygy2}
\end{center}
\end{table}

Note that many of the ratios $\frac{N_s}{N_s(\beta)}$ and $\frac{n+{\bar n}}{n_{\beta}+{\bar n}_{\beta}}$
in Table \ref{tab:SD13syzygy2} are identical, but as many are not: in particular only trival ratios
in the ``moth I - butterfly IV'' sub-sequence are identical. Perhaps more importantly, the ratios of 
$\frac{N_s}{N_s(\rm M_8)}$, which determine the ratios of periods $\frac{{\rm T}}{{\rm T}_{\rm M8}}$ , 
i.e., the slopes of linear sub-sequences are often not-as-good as with 
$\frac{n+{\bar n}}{n_{\rm M_8}+{\bar n}_{\rm M_8}}$: for example
for ``moth I'' orbit, the measured value $\frac{{\rm T}}{{\rm T}_{\rm M8}}=2.62031$ is closer to 
$\frac{n+{\bar n}}{n_{\rm M_8}+{\bar n}_{\rm M_8}} = 2.5$ than to $\frac{N_s}{N_s(\rm M_8)}=3$. 
We leave it to the interested reader to compare the quality of the linear fit in Fig.
\ref{fig:period_syzygy} with that in Fig. \ref{fig:period} to $N_s$ and to draw their own conclusions.

\begin{center}
\begin{figure}[h!]
\includegraphics[width=1.0\columnwidth]{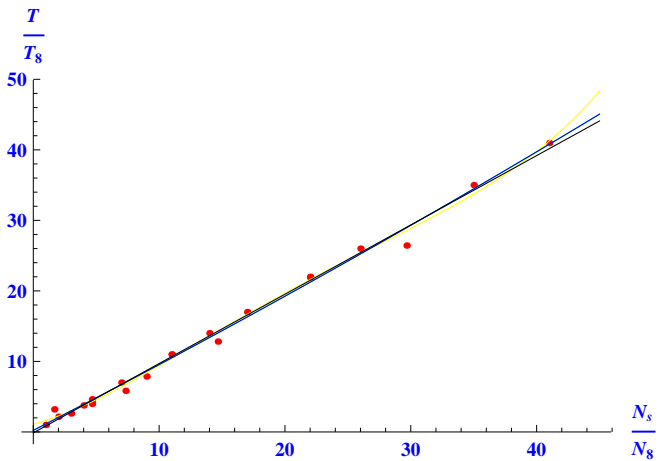}
\caption{(color on line) The rescaled periods $T_r(w)$ of presently known (zero-angular-momentum) 
three-body orbits divided by the period of the figure-8 orbit $T_r(w_8)$ 
versus the length $N_s$ of integer sequence divided by the length $N_8 = 6$ of the figure-eight orbit
integer sequence.
Four (linear, quadratic, cubic and quartic) fits are shown as solid lines of different colors.} 
\label{fig:period_syzygy}
\end{figure}
\end{center}


\begin{thebibliography}{9}

\bibitem{Landau}
L.D. Landau and E. M. Lifshitz, {\it Mechanics}, 
(3rd ed.) Butterworth-Heinemann,  Oxford (1976).

\bibitem{Montgomery1998}
R. Montgomery, 
Nonlinearity {\bf 11}, 363 - 376 (1998).

\bibitem{Suvakov:2013}
Milovan ~\v Suvakov, and V.~Dmitra\v sinovi\' c, 
Phys.\ Rev.\ Lett. {\bf 110}, 114301 (2013).

\bibitem{Suvakov:2014}
Milovan~\v Suvakov, and V.~Dmitra\v sinovi\' c, 
Am. J. Phys. {\bf 82}, 609 - 619 (2014).

\bibitem{Suvakov:2013b}
Milovan ~\v Suvakov, 
Celest. Mech. Dyn. Astron. {\bf 119}, 369-377 (2014). 
  
\bibitem{Shibayama}
Milovan ~\v Suvakov and Mitsuru Shibayama, 
``Three Topologicaly Nontrivial Choreographic Motions of Three Bodies'', 
submitted to Celest. Mech. Dyn. Astron., (2015).

\bibitem{Moore1993}
C. Moore, 
Phys. Rev. Lett. {\bf 70}, 3675 (1993).

\bibitem{Chenciner2000}
A. Chenciner, R. Montgomery,
Ann. Math. {\bf 152}, 881 (2000).

\bibitem{Simo2002}
C. Sim\'{o},
Contemp. Math. {\bf 292}, 209-228 (2002).

\bibitem{Galan2002}
J. Gal\' an, F. J. Mu\~noz-Almaraz, E. Freire, E. Doedel, and A. Vanderbauwhede, 
Phys. Rev. Lett. {\bf 88}, 241101 (2002).






\bibitem{Dmitrasinovic:2014lha} 
  V.~Dmitra\v sinovi\' c, M.~\v Suvakov and A.~Hudomal,
  Phys.\ Rev.\ Lett.\  {\bf 113}, 101102 (2014).

\bibitem{Kepler_chiba2015}
V.~Dmitra\v sinovi\' c and Milovan ~\v Suvakov, contribution to the Proceedings of 
``$46^{\rm rd}$ Symposium on Celestial Mechanics and N-Body Dynamics'', Chiba, Japan, 
November (2014).

\bibitem{Supplement}
Supplementary Information.

\bibitem{Henon1976}
M. Henon, 
Celest. Mech. {\bf 13}, 267 (1976).

\bibitem{Henon1977}
M. Henon, 
Celest. Mech. {\bf 15}, 243 (1977).


\bibitem{gallery}
http://suki.ipb.ac.rs/3body/index.php 

\bibitem{Moeckel:2014}
R. Moeckel and R. Montgomery, 
arXiv:1412.2263 (math.DS).

\bibitem{Heggie:2000ad} 
  Douglas C. Heggie,
  Mon.\ Not.\ Roy.\ Astron.\ Soc.\  {\bf 318}, L61 (2000)

\bibitem{Dubrovin}
B.A. Dubrovin, A.T. Fomenko and S.P. Novikov, 
{\it Modern Geometry - Methods and Applications (Part III. Introduction to Homology Theory)} 
(Springer, New York, 1990)

\bibitem{Vogtmann2002}
K. Vogtmann, Geometriae Dedicata {\bf 94}, 1-31 (2002).



\end{thebibliography}
\end{document}